\documentclass[trackchanges,twocolumn]{aastex7}
\usepackage{listings}
\usepackage{xcolor}

\begin{document}

\title{Residual 1D CNN for Low $\Sigma_{\mathrm{SFR}}$ Regression: A Design Note}

\author[orcid=0000-0001-8894-0854]{Po-Chieh Yu}
\affiliation{Taiwan Astronomical Research Alliance (TARA), Taiwan}
\altaffiliation{https://tara.tw}
\affiliation{Institute of Astronomy and Astrophysics, Academia Sinica, Taipei, Taiwan}
\email[show]{pcyu@asiaa.sinica.edu.tw}

\begin{abstract}
This technical note describes the design and modular implementation of a one-dimensional convolutional neural network (1D CNN) adapted from residual networks (ResNet), developed for photometric regression tasks with an emphasis on low star formation rate surface density ($\Sigma_{\mathrm{SFR}}$) inference. The model features residual block structures optimized for sparse targets, with optional loss weighting and diagnostic tools for analyzing residual behavior. The implementation (version \texttt{v1.4}) originated during a collaborative project and is documented here independently. No external data are reproduced or analyzed. This note provides a reusable architectural reference for scalar regression problems in astronomy and related domains.
\end{abstract}


\section{Introduction} 

Many regression tasks in astronomy, such as estimating star formation rates (SFRs) or stellar mass, require mapping from noisy features (e.g., broadband photometry) to scalar outputs. However, current models often suffer from underfitting or overfitting due to limited flexibility and weak residual control.

Predicting galaxy SFRs from photometric and redshift data is particularly challenging in large-scale extragalactic surveys. While convolutional neural networks (CNNs; \citealt{Lecun1998}) have shown good performance in various domains, their application to low SFR regimes, especially low surface densities ($\Sigma_{\mathrm{SFR}}$), remains unstable and underexplored.

Several factors contribute to this difficulty:
\begin{itemize}
\item \textbf{Data imbalance:} The long-tailed distribution of SFRs results in a small number of low-SFR systems in the training set, making it difficult to assign appropriate loss weights to these underrepresented samples \citep{Yang2021}.
            
    \item \textbf{Feature degeneracy and noise:} Photometric features of low-$\Sigma_{\mathrm{SFR}}$ galaxies often resemble those of passive ones. In such regimes, signal-to-noise ratios tend to be low, and limited photometric resolution blurs structural indicators, making it difficult for models to resolve the physical difference between quiescent and weakly star-forming systems \citep[see e.g.,][]{Pacifici2016}.
    
    \item \textbf{CNN limitations:} Standard one-dimensional (1D) CNNs primarily capture local patterns and may fail to integrate broader structural cues, especially when input features are sparse, noisy, or poorly aligned with spatial morphology. Similar limitations have been noted in time-series applications, such as light curve classification, where 1D CNNs alone struggle to capture long-range dependencies \citep{fawaz2019deep}.
\end{itemize}

In some studies, a chain architecture is used in which intermediate predictions, such as stellar mass, are fed into subsequent networks to predict additional parameters like SFR \citep{Alfonzo2024}. While this approach may improve accuracy in edge cases such as quenched galaxies, it introduces additional challenges for interpretation. Since the input includes model-derived quantities, it becomes difficult to separate physically meaningful correlations from artifacts produced by the prediction pipeline. This structure may also increase the risk of information leakage, especially if the model implicitly learns from its own prior outputs.

1D CNNs also face vanishing gradients under low-signal or flat loss conditions. Residual blocks, introduced in ResNet architectures \citep{He2016} help by adding skip connections that preserve gradient flow and activation across layers, supporting deeper networks that can still learn from weak signals.

While ResNet-based architectures have been applied in recent works using chain-style frameworks to predict multiple physical properties from galaxy images \citep{Alfonzo2024}, our approach differs in both design and motivation. This work presents a standalone residual 1D CNN designed to learn low-$\Sigma_{\mathrm{SFR}}$ patterns directly, without relying on additional regression targets. The architecture and training strategy are designed to ensure stable convergence and maintain physically meaningful gradients, even under low signal-to-noise conditions.

\section{Model Architecture} \label{sec:model}

The proposed model is a residual 1D CNN designed to capture patterns in structured input vectors, such as photometric features and redshifts, and map them to scalar targets in a regression setting. The architecture is inspired by ResNet and consists of modular components that enable flexible tuning and expansion.

The network begins with an input layer followed by a series of convolutional blocks. Each block includes 1D convolutional layers with ReLU activation, L2 regularization, and residual skip connections to improve gradient flow and preserve feature continuity. The model preserves input resolution across all convolutional layers by using \texttt{stride=1} and \texttt{padding='same'}, avoiding explicit downsampling or pooling operations.

The base architecture includes four residual blocks with progressively increasing filter sizes:
\begin{itemize}
    \item An initial convolutional layer with 64 filters and kernel size 3
    \item A first residual block mirroring the initial configuration
    \item A second block with a $1 \times 1$ convolution for dimension adjustment, followed by a 128-filter layer
    \item Third and fourth residual blocks extending to 256 and 512 filters, respectively
\end{itemize}

Each residual block includes a skip connection to preserve information flow, which is especially important for learning low-amplitude signals. After the final block, the feature maps are flattened and passed through two fully connected layers: one with 256 units using \texttt{tanh} activation, and another with 128 units using ReLU. The output layer consists of a single neuron with linear activation for scalar regression.

The model is designed to support customization through:
\begin{itemize}
    \item Adjustable depth and filter width, by modifying the number and size of residual blocks
    \item Configurable residual merge strategies (e.g., additive or concatenation), though the current version uses additive merging
    \item Optional dropout layers for regularization, which can be inserted around dense layers if needed
\end{itemize}

To support stable training, the framework includes:
\begin{itemize}
    \item Learning rate scheduling (e.g., ReduceLROnPlateau logic)
    \item Early stopping based on validation feedback
    \item Sample weighting mechanisms for imbalanced regression targets
\end{itemize}

In experimental pipelines, additional outputs may be inserted to monitor residual behavior during training. Diagnostic hooks and visualization tools can be incorporated to analyze underfitting or overfitting patterns across different target regimes.

The model implementation is modular and compatible with both synthetic and structured input arrays. It is intended as a reusable architecture for regression problems involving noisy or low-amplitude targets.

A complete summary of the residual 1D CNN architecture, including layer configurations, filter dimensions, and activation functions, is provided in Appendix~\ref{tab:arch} for reference.

\section{Implementation Overview} \label{sec:implementation}

The model is implemented in Python using the TensorFlow (Keras) API. The current version is tagged as \texttt{v1.4}, featuring a modular codebase designed for structural flexibility and rapid experimentation. It includes utilities for input normalization, sequence padding, and batch-level data handling. Parameters such as filter sizes, residual depth, and activation functions are centrally configurable, with optional components such as dropout and L2 regularization available for tuning.

While this note contains no observational data or proprietary results, the architecture and training logic were originally developed and tested during a collaborative project using internal inputs. This technical note was written independently to document the core architecture in a version that does not depend on specific datasets.

The residual design was first introduced in early 2025 during internal development efforts to stabilize regression performance in $\Sigma_{\mathrm{SFR}}$ regimes. Initial attempts using standard 1D CNNs often led to training that stopped improving or convergence toward median values, particularly for low-$\Sigma_{\mathrm{SFR}}$ regimes. The residual framework, incorporating skip connections and multi-scale filters, was introduced to reduce these issues and became the foundation of subsequent modeling iterations.

The current form of the model preserves that design intent, serving as a clean architectural baseline for regression tasks involving sparse signals or imbalanced scalar targets. It is intended as a reusable reference for future work, whether applied to SFR estimation or to other structurally sensitive physical quantities.

\section{Design Rationale and Model History}

The architecture described above was shaped by practical limitations encountered during early development in 2025. Initial designs used standard 1D CNN pipelines, which proved unstable in low-$\Sigma_{\mathrm{SFR}}$ regimes. They often led to early loss plateauing or convergence toward median predictions. These behaviors were largely due to weak signal gradients in the photometric inputs and the sparse representation of low-$\Sigma_{\mathrm{SFR}}$ regions.

To address this, the design shifted toward residual learning. By adding skip connections and using convolutional layers with different filter sizes, the model improved signal flow through deeper layers and reduced gradient problems during training. Unlike cascaded models that rely on additional predictions, this architecture supports direct end-to-end learning from raw photometric features.

The loss function was also adjusted to better handle the imbalanced distribution of $\Sigma_{\mathrm{SFR}}$. Mean squared error (MSE) often gave too little penalty for errors at the low end, so the model uses Huber loss together with optional sample weights to make training more stable and predictions easier to interpret.

Other CNN architectures were also tested during early development, including standard 1D CNNs without skip connections and direct adaptations of the ResNet framework for photometric inputs. While both showed reasonable convergence, the ResNet-style models were less stable in low-$\Sigma_{\mathrm{SFR}}$ regions and harder to tune for sparse scalar targets. In contrast, the customized residual CNN used here with tailored skip connections and adjustable depth proved more robust and easier to adapt for astronomy-focused regression tasks.

This residual 1D CNN design was first prototyped during a collaborative phase, with early validation using shared observational data. The current note revisits that version as an independent architectural reference, written without reliance on proprietary datasets. It preserves the original intent and design rationale contributed by the author, and is intended for reuse in other regression tasks involving sparse signals or long-tailed target distributions.

\section{Conclusion}

While many studies focus on predicting total SFR from photometry, direct regression of $\Sigma_{\mathrm{SFR}}$ remains underexplored. Compared to total SFR, which often correlates with global scaling features such as luminosity or redshift, $\Sigma_{\mathrm{SFR}}$ requires the model to resolve spatially encoded signatures, including morphological concentration, physical size, and color gradients. The present model addresses this gap by adopting a residual 1D CNN designed to handle the structural and statistical challenges of low-$\Sigma_{\mathrm{SFR}}$ inference. To our knowledge, no comparable architecture has been documented for this specific task.

This makes $\Sigma_{\mathrm{SFR}}$ prediction more sensitive to noise, feature degeneracy, and sample imbalance. It also places greater demands on the model architecture: local receptive fields must work together to capture meaningful global patterns. This motivates the use of residual connections, which help stabilize training when gradients are weak and allow deeper networks without losing important signals.

The impact of this architecture lies not in outperforming black-box pipelines on high-SNR targets, but in demonstrating that a task-specific residual 1D CNN can achieve stable learning and maintain interpretability in regimes traditionally difficult for CNNs.
This is especially relevant for normalized scalar targets that do not follow clear global trends or correlations.

While originally developed for low-$\Sigma_{\mathrm{SFR}}$ regression from photometric data, this residual 1D CNN architecture may be adapted to other scalar prediction tasks involving sparse, noisy, or imbalanced targets. Possible extensions include uncertainty estimation, auxiliary supervision, or attention-based modules. The modular implementation is intended to support such adaptation with minimal changes. Feedback or derivative implementations are welcome.

\begin{acknowledgments}
The author thanks the previous collaboration team for sharing the scientific background that informed the early stages of this model's development. While residual 1D CNNs have been applied in astronomy for tasks such as spectral classification and light curve analysis, this work presents a different version, designed specifically for scalar regression on photometric data. The model was developed independently and adjusted to address the challenges of low-$\Sigma_{\mathrm{SFR}}$ prediction.

P.-C.Y. is supported by the Taiwan Astronomical Research Alliance (TARA), with funding from the National Science and Technology Council (NSTC 113-2740-M-008-005). TARA is committed to advancing astronomy in Taiwan and laying the groundwork for a national observatory. This note is dedicated to the part of me that just wanted to build something right.
\end{acknowledgments}

\section*{Appendix 1: Architecture Summary}

The core architecture follows a residual convolutional stack inspired by ResNet, adapted for 1D photometric regression. Table~\ref{tab:arch} outlines the model structure used in \texttt{v1.4}.

\begin{table}[h!]
\centering
\caption{Residual 1D CNN Architecture}
\label{tab:arch}
\begin{tabular}{l l}
\hline
\textbf{Layer} & \textbf{Details} \\
\hline
Input          & 1D vector of photometric features \\
Conv1D         & 64 filters, kernel size 3, ReLU, L2 reg \\
Residual Block 1 & Conv1D(64) × 2, skip connection (identity) \\
Residual Block 2 & Conv1D(128), skip via Conv1D(1x1, 128) \\
Residual Block 3 & Conv1D(256), skip via Conv1D(1x1, 256) \\
Residual Block 4 & Conv1D(512), skip via Conv1D(1x1, 512) \\
Flatten         & -- \\
Dense           & 256 units, tanh, L2 reg \\
Dense           & 128 units, ReLU \\
Output          & 1 unit, linear activation (scalar regression) \\
\hline
\end{tabular}
\end{table}

In practice, the model is intended to be trained with Adam optimizer and callbacks such as 
ReduceLROnPlateau and EarlyStopping. While not applied to proprietary data in this note, 
the architecture is compatible with imbalanced regression tasks and sparse photometric inputs.

\bibliography{sample7}{}
\bibliographystyle{aasjournalv7}

\end{document}